\documentclass[preprint,pra,aps]{revtex4}
\usepackage{graphicx}
\usepackage{bm}

\def\ie{i.e.\ }
\def\PD#1#2{\frac{\partial #1}{\partial #2}}    
\newcommand\HAH {\hat{H}}

\def\k #1{|#1\rangle}
\def\b #1{\langle #1|}
\def\MR {\mathcal{R}}

\def\mrm #1{\mathrm{#1}}

      \def\hh2{$\mathrm{H+H_2}$}
      \def\dd2{$\mathrm{D+D_2}$}
      \def\tt2{$\mathrm{T+T_2}$}
\input epsf

\begin{document}
\title{Analytic solution to $N$ vs. $M$ photon phase control in an open two-level system}
\author{Heekyung Han and Paul Brumer }
 \affiliation{ Chemical Physics Theory Group,\\Department of Chemistry,
 and \\Center for Quantum Information and Quantum Control
\\University of Toronto\\ Toronto, Canada  M5S 3H6}

\date{\today}
\begin{abstract}
Decoherence effects on the traditional $N$ vs. $M$ photon coherent 
control of a two-level system are 
investigated, with 1 vs. 3 used as a specific example.
The problem reduces to that of a 
two-level system interacting with a single mode field, but with an 
effective Rabi frequency that depends upon the fundamental and third 
harmonic fields. The resultant 
analytic control solution is explored for a variety of parameters, with
emphasis on the dependence of control on the relative phase of the lasers.
The generalization to off-resonant cases is noted.
\end{abstract}
\maketitle
\vspace{0.2in}
\noindent
 \clearpage
\section{Introduction}\label{4-1}
Coherent control of  atomic and molecular dynamics using optical 
fields has attracted much attention, both theoretically and 
experimentally \cite{brumer_shapiro20,tannor_rice0,rabitz0}. Thus 
far, most theoretical  work has focused on the idealized case of  
isolated systems, where loss of quantum phase information due to 
decoherence, i.e. coupling to the environment, is ignored.  Such 
effects are, however, crucial to control in realistic systems, 
since loss of phase information results in loss of control.  For 
this reason efforts to understand control in external
environments \cite{chirped2}-- \cite{fainberg} and to compensate 
for the resultant decoherence (e.g., \cite{shor}--\cite{tannor0}) 
are of great interest.

There exist a number of basic interference schemes\cite{brumer_shapiro20} 
that embody the
essence of coherent control. One is the $N$ vs. $M$ photon
scenario where control results from interference between state
excitation using $N$ and $M$ photons simultaneously.
In this letter we provide an analytic solution for control in the
two-level $N$ vs. $M$ photon control scenario in the presence of 
decoherence. For simplicity, we examine the 1 vs. 3 photon case,
although the solutions obtained below apply equally well to the 
$N$ vs. $M$ photon case, with obvious changes in the input Rabi
frequencies and relative laser phases.

In 1 vs. 3 photon control\cite{brumer_shapiro1+3} a 
continuous wave electromagnetic
field composed of a superposition of a fundamental and third harmonic 
wave is incident on a system. By varying the relative phase and
amplitude of the fundamental and the third harmonic one can alter the
population of the state excited by the incident field. 
Clearly, decoherence can be expected to
diminish the 1 vs. 3 photon induced interference, 
and hence the control over excitation. Although extensive theoretical 
\cite{brumer_shapiro1+3} - \cite{LambroPra} and 
experimental \cite{elliott} - \cite{review_1+3} studies have been carried out 
on the 1 vs. 3 photon coherent control scenario, there has been no 
serious examination of the stability of this control scheme in an 
external environment, barring a derivation of a simple analytical 
expression for the autoionization of a two-level atomic system 
for  weak laser intensities, using the rate approximation 
\cite{Lambro}. Amongst the various possible influences of an environment on a 
system we focus on the loss of phase coherence, that is, dephasing. 
Dephasing is expected to occur on a time scale more relevant to 
control, since the duration of control field can be on the order of a 
picosecond or less, wheras the typical time scale for 
energy transfer is considerably longer \cite{energy_transfer,et2}.

In this paper we show that the 1 vs. 3 photon phase
control scenario (which controls the  population)
in a two-level system, when coupled to an 
environment, reduces to the analytically soluble 
monochromatic field case, but with an 
effective Rabi frequency that is determined by the relative phase and 
amplitudes of the two fields.  Sample results for control as a function
of relative laser phase in the 
presence of dephasing are then provided. The possiblity of solving the 
off-resonance case is also noted.

\section{1 + 3 Photon Control}
\subsection{Formalism}

Consider a two-level bound system interacting with an continuous wave (CW)
electromagnetic field and assume that the energy levels undergo  random Stark 
shifts without a change of state during collisions with an 
external bath, e.g.,  elastic collisions between atoms in a 
gas.  The CW field $E(t)$ is treated classically,
and the ground and the excited energy 
eigenstates  states, of energy $E_1$ and $E_2$ are denoted  
$\k 1$ and $\k 2$, respectively.

In general, the system density operator $\rho$  obeys the Liouville equation,
\begin{eqnarray}
\PD {\rho}{t}=-\frac {i}{\hbar}[\HAH(t),\rho]-{\MR}\rho.
\label{liouville12l}\end{eqnarray} Here $\HAH
(t)=\HAH_{\mrm{atom}}+\HAH_{\mathrm{int}}$, where the free atomic 
Hamiltonian term is 
\begin{equation}
\HAH_{\mrm{atom}}=E_1 \k 1 \b 1 + E_2 \k 2 \b 2
\end{equation}
and the atom-field interaction term within the dipole approximation is 
\begin{equation}
\HAH_{\mathrm{int}}= 
-E(t)[~\langle 1|d|2\rangle |1\rangle\langle 2| + 
\langle 2|d|1\rangle |2\rangle\langle 1|~ ]
\end{equation}
with electric dipole operator $d$.  The second 
term in Eq. (\ref{liouville12l}),  $\mathcal{R}$, is a dissipative term that can 
have a variety of nonequivalent forms associated with various 
master equations.  Below we assume simple exponential dephasing 
of the off-diagonal $\rho_{ij}$.

In the  simplest 1 vs. 3 control scenario, a two-level  system is subject to the  
linearly polarized laser field:
 \begin{eqnarray}
 E(t) &=&\frac{1}{2}[{\mathcal{E}}_f e^{i\omega_f t }e^{-i\phi_f}+{\mathcal{E}}_h e^{i\omega_h
 t}e^{-i\phi_h}+c.c.],
 \label{efield2l}
  \end{eqnarray}
where ${\mathcal{E}}_j$ is the real time-independent amplitude 
and $\phi_j$ is the phase of the corresponding field, with 
$j=h,f$. Here the subscripts $f,h$ denotes the fundamental and 
its third harmonic, and $``c.c."$ denotes the complex
conjugate of the terms that precede it. The fields have frequencies 
$\omega_f$ and $\omega_h=3\omega_f$, 
chosen so that the third-harmonic and the three fundamental photons are on 
resonance with the transition from the ground state $\k 1$ to the 
excited state $\k 2$.  In the standard scenario 
\cite{brumer_shapiro20,tannor_rice0,brumer_shapiro1+3}, control 
is obtained by changing the relative phase and amplitudes of two fields,  which 
results in the alteration of the degree of interference between 
the two pathways to the excited state.

Within the rotating-wave approximation, the slowly varying 
density-matrix elements of the states $\k 1$ and $\k 2$, 
$\sigma_{ii}=\rho_{ii}$, ($i=1,2$) and $\sigma_{21}=\rho_{21} 
e^{3i(\omega_f  t+ \phi_f)}$  obey the following set of equations:
\begin{eqnarray}
\frac{\partial \sigma_{11}}{\partial t} & = & - \textrm{Im}[(
\mu^{(3)}_{12} \mathcal{E}^3_f/\hbar + \mu_{12} \mathcal{E}_h 
e^{i\phi}/\hbar)\sigma_{21}]\nonumber\\
\PD {\sigma_{22}}{t}&=&
    {\mrm{Im}}[ ( {\mu}_{12}^{(3)}{\mathcal{E}}_f ^3 /\hbar+ {\mu}_{12}{\mathcal{E}}_h e^{i\phi}
    /\hbar)\sigma_{21}],
    \label{density_matrix112l}\nonumber\\
\PD {\sigma_{21}}{t}&=&-\gamma_{{p}} \sigma_{21} +\frac{i}{2} (
{\mu}_{21}^{(3)}{\mathcal{E}}_f ^3 /\hbar+ {\mu}_{21}{\mathcal{E}}_h
e^{-i\phi}
    /\hbar)(\sigma_{11}-\sigma_{22}),
 \label{density_matrix212l}\end{eqnarray}
with
\begin{eqnarray}
\mu_{12}^{(3)}\equiv\frac{1}{(2\hbar)^2}\sum_{n,m}\frac{\mu_{1n}\mu_{nm}\mu_{m2}}
{(\omega_{n1}-\omega_f)(\omega_{f}-\omega_{2m})}\,\,.
\label{mu122l}\end{eqnarray} Here  $\gamma_p$ is the dephasing rate,
$\omega_{nm}$ is the 
frequency difference between levels $\k n$ and $\k m$ and 
$\mu_{nm}\equiv\b n d \k m$. The quantities $\mu_{12}$  and 
${\mu}_{12}^{(3)}$ denote the one-photon matrix element for the 
harmonic field and the effective three-photon matrix element  for 
the fundamental field for the  $\k 1 
\rightarrow \k 2$ transition. Below, we use ${\mu \equiv \mu_{12}}$ and 
$\mu^{(3)} \equiv \mu^{(3)}_{12}$, 
omitting the subscripts for simplicity. The controllable relative 
phase is $\phi = \phi_{h}-3\phi_{f}$.

It is convenient to define the one- and three-photon Rabi frequencies 
by
$\Omega_{h}=\mu{\mathcal{E}}_h/\hbar$ and 
$\Omega_{f}=\mu^{(3)}{\mathcal{E}}_f^3/\hbar,$
given  in terms of their amplitudes and phases, by
$\Omega_{h}=|\Omega_{h}|e^{i\theta_{h}}$ and
$\Omega_{f}=|\Omega_{f}|e^{i\theta_f}$.  Note that, although 
$\mu$ and $\mu^{(3)}$ are real for a bound system, we derive all 
the equations under the assumption that they can be complex so that
the analysis can be extended to complex matrix elements 
arising in transitions to the continuum. Since ${\mathcal{E}}_h$ 
and ${\mathcal{E}}_f$ are real and positive, $\theta_h$ and 
$\theta_f$ are determined by $\mu$ and $\mu^{(3)}$:
\begin{eqnarray}
e^{i\theta_{h}}&=&\mu/|\mu|,\label{mu1}\\
e^{i\theta_f}&=&\mu^{(3)}/|\mu^{(3)}|. \label{mu3}\end{eqnarray}
To amalgamate these Rabi frequencies and the relative laser 
phase of $\phi$, we define the effective Rabi frequency 
$\Omega_{\rm{eff}}$:
\begin{eqnarray}
\Omega_{\rm{eff}} e^{i\theta} \equiv \Omega_h e^{i\phi} + 
\Omega_f = | \Omega_h e^{i\phi} + \Omega_f | e^{i\theta},
\label{omegaeff}\end{eqnarray}
where $\Omega_{\rm{eff}}$ is real and positive.   Here 
$\Omega_{\rm{eff}}$ and $\theta$ are related to  $\Omega_h$ and 
$\Omega_f$ as
\begin{eqnarray}\Omega_{\rm{eff}} & = & \sqrt{|\Omega_h| ^2 + |\Omega_f |^2 +2|\Omega_h
\Omega_f| \cos{\Phi}}\label{omegaeff3}, \\
\tan{\theta} & = &
[{\sin{(\phi+\theta_h)}+\frac{|\Omega_f|}{|\Omega_h|}\sin{\theta_f}}]/[ 
{\cos{(\phi+\theta_h)}+\frac{|\Omega_f|}{|\Omega_h|}\cos{\theta_f}}],
\end{eqnarray} 
where 
\begin{eqnarray}
\Phi=\phi+\theta_h-\theta_f. \label{Phid}\end{eqnarray}

It is worth noting some  features of $\Omega_{\rm{eff}}$ that are evident
from Eq. (\ref{omegaeff3}). First,  
the total excitation probability obtained in 
lowest order perturbation theory for 1 vs. 3 photon phase control in 
a two-level system \cite{brumer_shapiro1+3} is 
proportional to $\Omega_{\rm{eff}}$.
Hence, $\Omega_{\rm{eff}}$ can be used to predict the 
controlled population, and its dependence on $\phi$, when the fields
are weak. Further (see below), $\Omega_{\rm{eff}}$ 
plays a major role in determining the transient behavior of the 
excited state population for any  field intensity  in the 
absence or presence of dephasing. Second, the interference 
term in Eq. (\ref{omegaeff3}) can be  controlled by varying 
$\Phi$, that is, by manipulating  the relative phase $\phi$ of the 
two fields. Since $\mu$ and $\mu^{(3)}$ are real in a bound 
system, possible values of $\theta_h-\theta_f$ are 0 and 
$\pm\pi$. When $\theta_h-\theta_f$=0, \ie, $\mu \mu^{(3)}>0$, 
$\cos\Phi = \cos\phi$. On the other hand, when 
$\theta_h-\theta_f=\pm\pi$, \ie, $\mu \mu^{(3)}<0$, $\cos\Phi = 
-\cos\phi$. Thus, opposite interference effects are observed 
depending on the signs of $\mu$ and $\mu^{(3)}$.  Third,  
$||\Omega_h|-|\Omega_f||\leq \Omega_{\rm{eff}} \leq 
|\Omega_h|+|\Omega_f|$ so that maximal  interference effects occur 
when  $|\Omega_h|=|\Omega_f|$. If $|\Omega_h|\neq|\Omega_f|$, the 
smallest $\Omega_{\rm{eff}}$ is not zero, and thus complete 
destructive interference, that is, zero excitation from the ground to 
the excited state, does not occur.
	      
Rewriting Eq. (\ref{density_matrix212l}) in terms of
$\Omega_{\rm{eff}}$ and  $\theta$,
\begin{eqnarray}
\frac{\partial \sigma_{11}}{\partial t} & = & 
-\textrm{Im}[\Omega_{\rm{eff}} 
e^{i\theta}\sigma_{21}] \\
 \PD {\sigma_{22}}{t}&=&
    {\mrm{Im}}[ \Omega_{\rm{eff}} e^{i\theta} \sigma_{21}],
    \label{redensity_matrix112l}\\
\PD {\sigma_{21}}{t}&=&-\gamma_{{p}} \sigma_{21} +\frac{i}{2} [
{\Omega_{\rm{eff}}}e^{-i\theta} ](\sigma_{11}-\sigma_{22}).
 \label{redensity_matrix212l}\end{eqnarray}
and introducing $u=2\,{\mrm{Re}}(\sigma_{12}e^{-i\theta})$,
$v=2\,{\mrm{Im}}(\sigma_{12}e^{-i\theta})$,  and 
$w =\sigma_{22} -\sigma_{11}$, gives
\begin{eqnarray}
du/dt=-\gamma_{{p}}u, \label{cut}\\
dv/dt=-\gamma_{{p}}v+\Omega_{\rm{eff}} w, \label{cvt}\\
dw/dt=-\Omega_{\rm{eff}}v.\label{cwt}
\end{eqnarray}

The resultant equations are now of standard form\cite{torrey}, but with
$\Omega_{\rm{eff}}$ replacing the Rabi frequency of the single field case
discussed in Ref \cite{torrey}.
Note that the longtime steady-state solution to Eqs. (\ref{cut}) to 
(\ref{cwt}) is found by setting $du/dt=dv/dt=dw/dt=$0, giving  
$u(t\rightarrow\infty)=v(t\rightarrow\infty)=w(t\rightarrow\infty)=$0. 
This implies that, regardless of initial conditions and for 
sufficiently large time, pure dephasing leads to an equilibrium 
state  with equal populations in the ground and the excited 
states and with no remaining coherence.

Substituting Eq. (\ref{cvt}) into Eq.
(\ref{cwt}) gives a simple equation for $w$ :
\begin{eqnarray}
d^2 w/dt^2+\gamma_{{p}}dw/dt+\Omega_{\rm{eff}}^2 w=0.\label{w2t}
\end{eqnarray} 
In the important case where
initially the ground state is populated and the  coherence is 
zero [i.e., $w(0)=-1$, $u(0)=v(0)=0$], the 
excited state population $\rho_{22}= \sigma_{22}$ is given by
\begin{eqnarray}
\rho_{22} =-\frac{e^{-\frac{\gamma_{{p}}
t}{2}}}{2}[\cos(st)+\frac{\gamma_{{p}}}{2s}\sin(st)]+\frac{1}{2}\,\,
&{\mrm{for}}& \gamma_{\mrm{p}}< 2\Omega_{\rm{eff}},   \label{oscillation} \\
 \rho_{22}=\frac{[-\lambda_2 e^{\lambda_1 t}+\lambda_1 e^{\lambda_2 t} ]}{2(\lambda_2-\lambda_1)}+\frac{1}{2}\,\,\,\,\,\,\,\,\,\,\,\,
&{\mrm{for}}& \gamma_{{p}}> 2\Omega_{\rm{eff}}, \label{mono}\\
  \rho_{22}=-\frac{e^{-\frac{\gamma_{{p}}
t}{2}}}{2}(1+\frac{\gamma_{{p}}t}{2})+\frac{1}{2}\,\,\,\,\,\,\,\,\,\,\,\,
&{\mrm{for}}& \gamma_{{p}}= 2\Omega_{\rm{eff}}, \label{expo}
\end{eqnarray}
where  $s=\frac{1}{2}\sqrt{4\Omega_{\rm{eff}}^2-\gamma_{{p}}^2}$, and
$\lambda_{1,2}=\frac{1}{2}[-\gamma_{{p}}\pm\sqrt{\gamma_{{p}}^2-4\Omega_{\rm{eff}}^2}]$. 
 The general behavior of the 
solution is seen to be determined by relative size of the 
dephasing time and the period of the Rabi oscillation. Analogous 
analytic results can be obtained for 
 $\sigma_{12}$ which decays with rate $\gamma_{\rm{p}}$.  
 If the external field is intense enough so that $\gamma_{{p}}<2\Omega_{\rm{eff}}$, then $ \rho_{22}$ shows  oscillations that are exponentially 
damped  with time.  On the other hand, if dephasing 
 dominates over the Rabi oscillation, so that $\gamma_{{p}}> 
2\Omega_{\rm{eff}}$ or $\gamma_{{p}}= 2\Omega_{\rm{eff}}$, $\rho_{22}$ increases 
monotonically. However, in all cases $ \rho_{22}$ reaches a 
stationary value of 0.5 at long times and  $\rho_{22}\sim 
\frac{\Omega_{\rm{eff}}^2  t^2}{4}$ for  short times.

\subsection{Sample Computations}

\begin{figure}[ht]
    \includegraphics[angle=0,scale=1.]{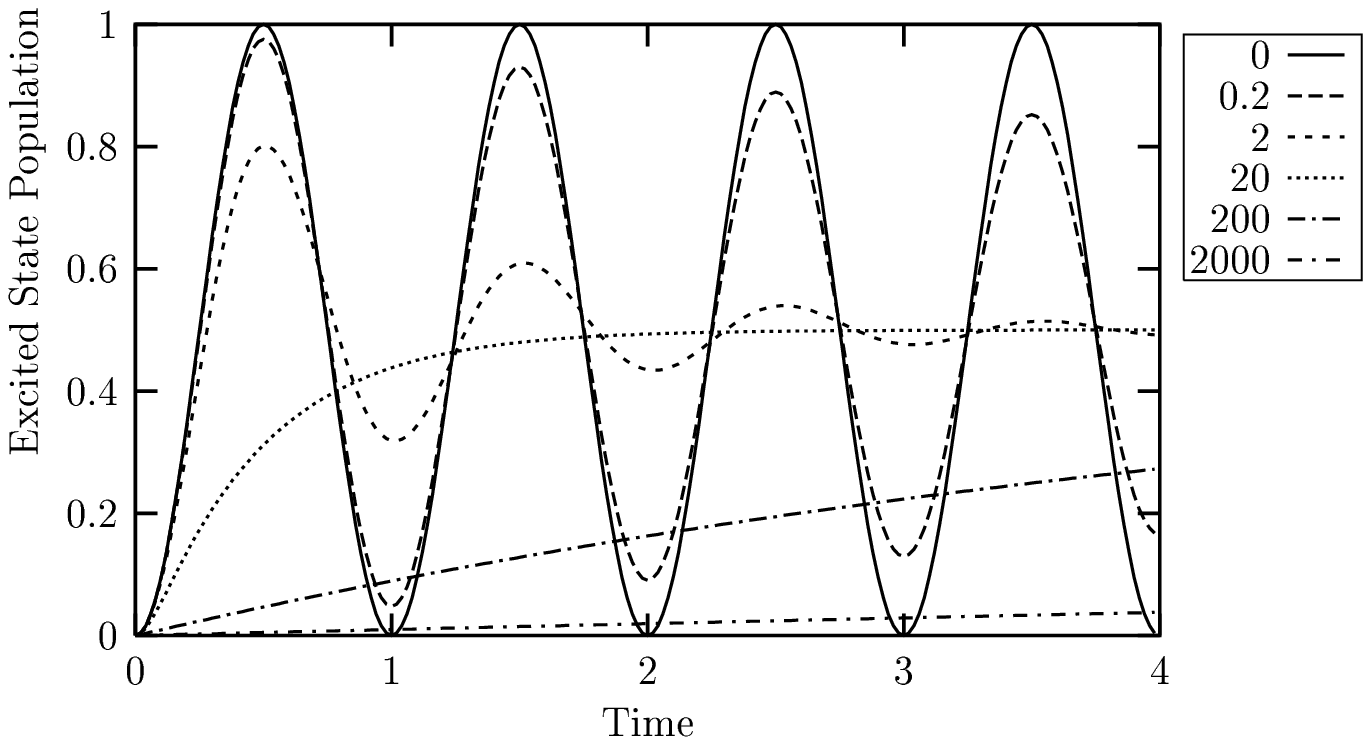}
    \caption{}
   \label{2lcwanal}
   \end{figure}

The behavior of the excited state population for several values of
$\gamma_{{p}}$  for a given value of $\Omega_{\rm{eff}}$ (here chosen as 2$\pi$)
is sketched in Fig. \ref{2lcwanal}.
For $\gamma_{{p}}<\,2\Omega_{\rm{eff}}$, the introduction of dephasing 
increases the period of the oscillation and causes the amplitudes 
to decay as $e^{-\frac{\gamma_{{p}} t}{2}}$. Although this is a CW laser 
field case, we can extract the result for the field being 
switched off at a specific time, i.e., a square pulsed laser 
which is  on from $t=0$ to $t=t_f$, by examining the population 
at time $t_f$. (This assumes that there are no additional energy levels
excited by the frequency breadth of the truncated CW source).
Significantly, one can  end up with an increased 
$\rho_{22}$ even for a larger dephasing, depending on the pulse 
duration. For example, assume that we turn off the field at 
$t$=1. If there is no dephasing, then the excited state 
population at $t$ = 1 is 0  and is thus less than that of any of 
the other cases with dephasing.
On the other hand, for  $\gamma_{{p}}>\,2\Omega_{\rm{eff}}$ (here
$\gamma_{{p}} > 4\pi$),  there is no oscillation; 
$\rho_{22}$ just increases monotonically towards 0.5, where the 
system reaches the steady-state slower with increasing  
dephasing. If we were to consider a pulse rather than a CW laser field for 
this relatively strong dephasing 
case, the excited state population would be expected to increase  up to 
0.5 with the increase in the pulse duration.

Typical behavior of $\rho_{22}$ and of the 1 vs. 3 photon phase control 
profile (i.e., $\rho_{22}$ as a function of generic phase control 
variable $\Phi$) for several 
values of Rabi frequencies and $\gamma_{{p}}$ are shown in 
Fig. \ref{control2lt}.  Here we assume that  the fields are 
abruptly turned off at the times indicated in the figure captions 
to produce a square pulse and the intensities are chosen so that 
$|\Omega_h | = | \Omega_f |$, to enhance the interference effects.
The effective Rabi frequency is 
then $\Omega_{\rm{eff}} = |\Omega_{h}|~\sqrt{2(1+\cos{\Phi})}$.  
While $\Phi=0$ leads to a complete constructive interference of 
the two transition amplitudes, $\Phi=\pi$ leads to a complete 
destructive interference, \ie, no excitation from the ground to 
the excited state. 

The typical control behavior seen in Fig. \ref{control2lt} depends upon 
the pulse duration, as well as upon $\Omega_{\rm{eff}}$ and 
$\gamma_{p}$.  For example, when the field is weak and 
$\gamma_{p} = 0$, then $\rho_{22}$ is 
given by [Eq (\ref{oscillation})-(\ref{expo})]  
\begin{equation} \rho_{22} =  \frac{1}{2}[1-\cos(\Omega_{\rm{eff}} 
t)]~~{\mrm{for}}~~\gamma_{{p}} = 0 \label{precos}  
\end{equation}
\begin{equation} \approx \frac{1}{2} |\Omega_h|^2 (1+\cos 
\Phi)t^2~~;~~ {\mrm{for}}~ \Omega_h~{\mrm{small}}. \label{cos} \end{equation} 
Hence, the system shows a ``$\cos \Phi$ rule". [ A similar rule obtains from
Eq. (\ref{oscillation}) - Eq. (\ref{expo}) when 
$\gamma_{\mrm{p}}< 2\Omega_{\rm{eff}}$ and $st << 1 $, when 
$\gamma_{\mrm{p}}> 2\Omega_{\rm{eff}}$ and $\lambda_1 t$ and $\lambda_2 t$ are 
much less than one, and for $\gamma_{\mrm{p}}= 2\Omega_{\rm{eff}}$ when
$\gamma_p t/2<<1$.] Note also that Eq. (\ref{precos}) predicts oscillatory
behavior of $\rho_{22}$ as a function of $\Omega_{\rm{eff}}$ at fixed $t$,
as observed later below.

The control 
profiles for small $\Omega_h$ and $\gamma_p=0$ (thin dashed lines in Fig.
\ref{control2lt})  are then seen to be 
monotonically decreasing from the maximum excitation at $\Phi=0$, 
to zero excitation at $\Phi=\pi$, i.e. they follow the ``$\cos \Phi$ rule".
By contrast, for  strong 
intensity (thin solid lines in Fig. \ref{control2lt})  in which 
there are  many Rabi cycles during the pulse, the control curve 
is not necessarily monotonic since the final excited 
populations are determined by the time at which the fields are 
turned off. Introducing dephasing is seen to lead to a decreased range 
of control whose magnitude depends on the relative strength of 
the dephasing and on the effective Rabi frequencies, according to 
Eqs. (\ref{oscillation}) - (\ref{expo}).

 Figure \ref{control2lt} 
demonstrates that phase control profiles are strongly dependent 
on the pulse duration.
For  weak intensities, as the pulse duration increases, 
the degree of control improves and the control curve continues to 
approximately follow a $\cos{\Phi}$ law (e.g., Eq. (\ref{cos})). This 
behavior is seen  both in the absence and in the presence of 
dephasing, although dephasing reduces the yield for a given pulse 
duration. In the strong field case the control 
profile  varies strongly with pulse duration. In particular, 
with $\gamma_{{p}} = 0$,  if the pulse duration is smaller 
than the oscillation period (=1/2) of the 
$\Omega_{\rm{eff}}(\Phi=0)$ case, then $\rho_{22}$ decreases 
with increasing $\Phi$, as shown in Fig. \ref{control2lt}(a).  
For the pulse duration greater than that period,  
the control profile
 no longer follows $\cos{\Phi}$ and the maximal yields start to appear at
$\Phi\neq 0$, as shown in Figs. \ref{control2lt}(b) to (d).   
 In all strong intensity cases, the addition of 
dephasing results in a decay of $\rho_{22}$ with a rate of 
$e^{-\frac{\gamma_{{p}}t}{2}}$ for a given 
$\Omega_{\rm{eff}}$.  Thus the degree of the control worsens in 
the presence of dephasing as the pulse duration increases.  Note that 
the introduction of dephasing 
leads to a degree of control $C$ that converges to 0.5,
where $C$ is defined as the 
difference between the maximum and minimum excited state populations.

Finally, we note that this treatment can be extended in two directions. 
The most obvious is to extend it to the general two-level
$N$ photon + $M$ photon
interference scenario\cite{brumer_shapiro20} where the structure of 
the problem is exactly the same as that of the 1 vs. 3 photon case.
The equations above therefore hold, but with the one and three photon
Rabi frequencies and phases replaced by the $N$ and $M$ photon Rabi
frequencies and phases.
The second is to consider the more general case that includes 
the equal detuning of both fields from the $\k 1\rightarrow \k 2$ 
transition, i.e. $\delta=\omega_{21}-\omega_h=\omega_{21}-3\omega_f$, 
and where the populations of  levels $\k 2$ and $\k 1$ decay with 
the same rate  $\gamma_d$. Then Eqs. (\ref{redensity_matrix112l}) 
to (\ref{redensity_matrix212l}) become:
\begin{eqnarray}
\PD {\sigma_{11}}{t}&=&-\gamma_d(\sigma_{11}-\sigma_{1e})
    -{\mrm{Im}}[ \Omega_{\rm{eff}} e^{i\theta} \sigma_{21}],
    \label{gredensity_matrix112l}\\
\PD {\sigma_{22}}{t}&=&  -\gamma_d(\sigma_{22}-\sigma_{2e}) +{\mrm{Im}}[
\Omega_{\rm{eff}}e^{i\theta} \sigma_{21}], \label{gredensity_matrix222l}\\
\PD {\sigma_{21}}{t}&=& -(\gamma_d+\gamma_{{p}}+i\delta)\sigma_{21}
+\frac{i}{2} [ {\Omega_{\rm{eff}}}e^{-i\theta}
](\sigma_{11}-\sigma_{22}).
 \label{gredensity_matrix212l}\end{eqnarray}
Here $\sigma_{1e}$ and $\sigma_{2e}$ are the steady-state values 
of $\sigma_{11}$ and $\sigma_{22}$, respectively, when 
$\Omega_{\rm{eff}}=0$, and are introduced  to allow for 
relaxation to equilibrium. In terms of $u$, $v$, and $w$, 
the above equations lead to
\begin{eqnarray}
du/dt= -\delta v-\frac{ u}{T_2}, \label{gcut}\\
dv/dt=\delta u -\frac{ v}{T_2}+\Omega_{\rm{eff}} w, \label{gcvt}\\
dw/dt=-\frac{ (w-w_e)}{T_1}+\Omega_{\rm{eff}}v.\label{gcwt}
\end{eqnarray}
where $w_e=\sigma_{2e}-\sigma_{1e}$,  $T_1=1/\gamma_d$ and
$T_2=1/{(\gamma_d+\gamma_{{p}})}$. Note that these are then of the 
same form as the usual Bloch equations for a monochromatic field. 
Torrey gave detailed analytical solutions for these equations in the
monochromatic field case \cite{torrey, eberly} and the same 
analytical solutions for the 1 vs. 3 photon phase control case can be
used, where the single field $\Omega$ considered by Torrey 
is replaced by $\Omega_{\rm{eff}}$. We do not pursue this direction 
in this letter.

\section{Summary}

In summary, we have obtained an analytic solution for $N$ vs. $M$ 
photon phase control of a two-level system in an environment 
described by a $1/\gamma_{{p}}$ dephasing time, with $N=1$ and $M=3$
as a specific example.  The results should serve as a prototype for 
understanding the 
results of  $N$ vs. $M$ photon phase control in more complicated systems,
such as controlled Xenon ionization and IBr photodissociation\cite{elsewhere}. 

\vspace{0.2in}
{\bf Acknowledgements} This work was partially supported by Photonics
Research Ontario and by the Natural Sciences and Engineering Research
Council of Canada

\newpage
Figure 1.  Excited state population as a function of time for 
various dephasing rates, $\gamma_{{p}}$ shown inside the box 
for $\Omega_{\rm{eff}}=\,2\pi$. All the variables are in dimensionless 
units. Data points are connected by straight lines as a guide.
\vspace{0.5in}

Figure 2.  Excited state population versus relative phase for two 
different $|\Omega_h|$: Solid lines and dashed lines denote the 
case at $|\Omega_h|=2\pi$ and $|\Omega_h|=\pi/5$, respectively. 
Thin lines and thick lines denote the case at 
$\gamma_{{p}}=0$ and $\gamma_{{p}}=\pi$, respectively. 
Fields are turned off at (a) $t=0.25$, (b) $t=0.25\times 2$, (c) 
$t=0.25\times 3$, and (d) $t = 0.25 \times 8$. Note that the data
in panel (d) is too widely spaced to produce the last $\rho_{22}=1$ 
maxima, whose exact locations can be predicted from Eq. (\ref{cos}).
All the variables are in dimensionless units. Data points are connected
by straight lines as a guide.

\clearpage
\clearpage

\begin{figure}[htbp]				        
\begin{center}
   \includegraphics[angle=0,scale=1.]{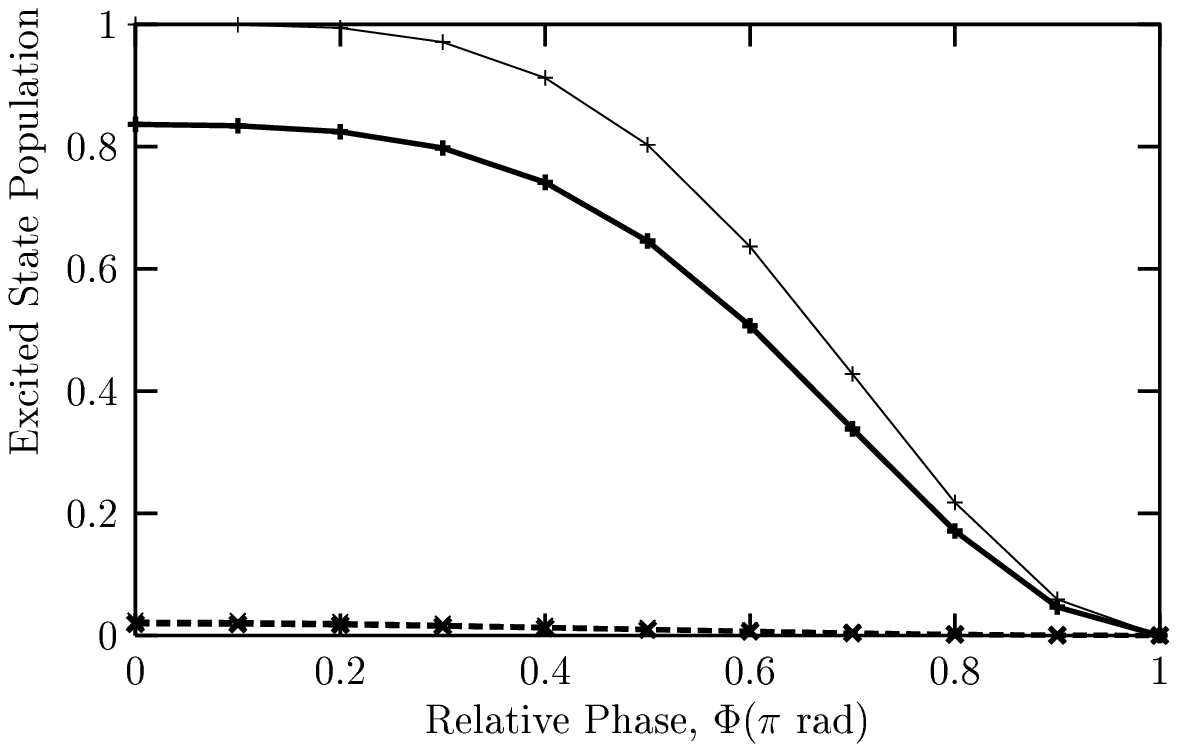}\\
   \includegraphics[angle=0,scale=1.]{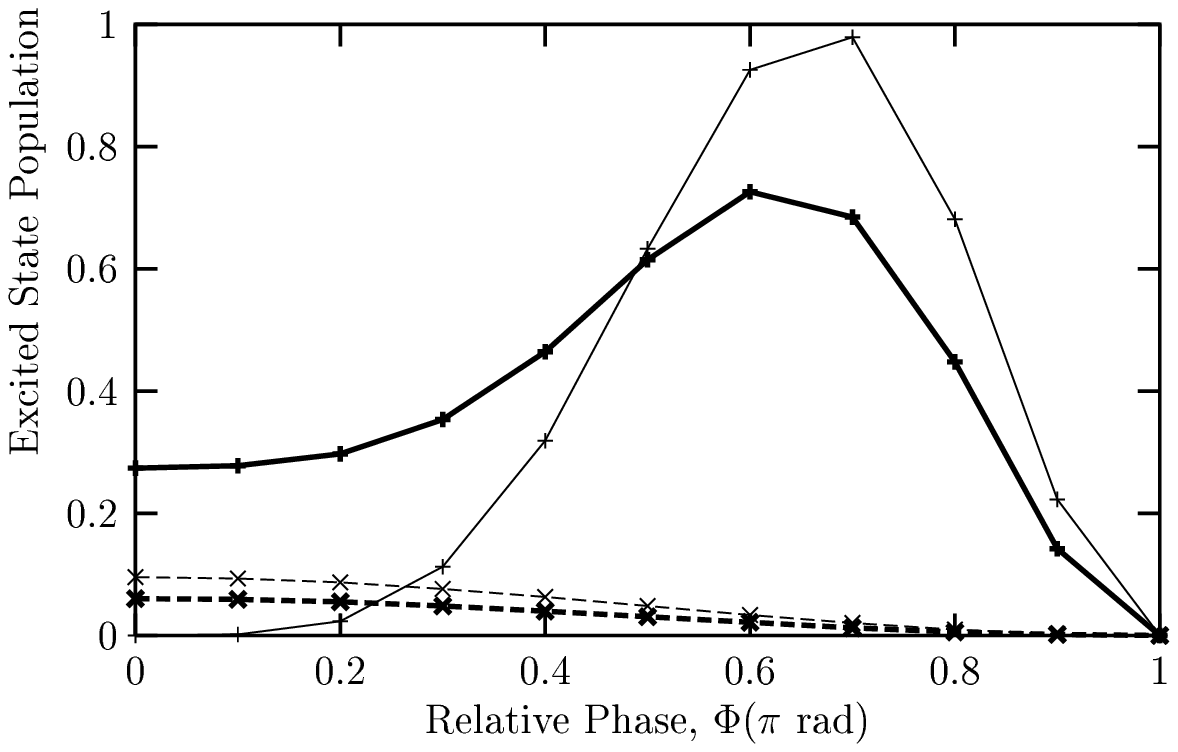}\\
  \end{center}\end{figure}
   \clearpage
\begin{figure}[htbp]				        
\begin{center}
   \includegraphics[angle=0,scale=1.]{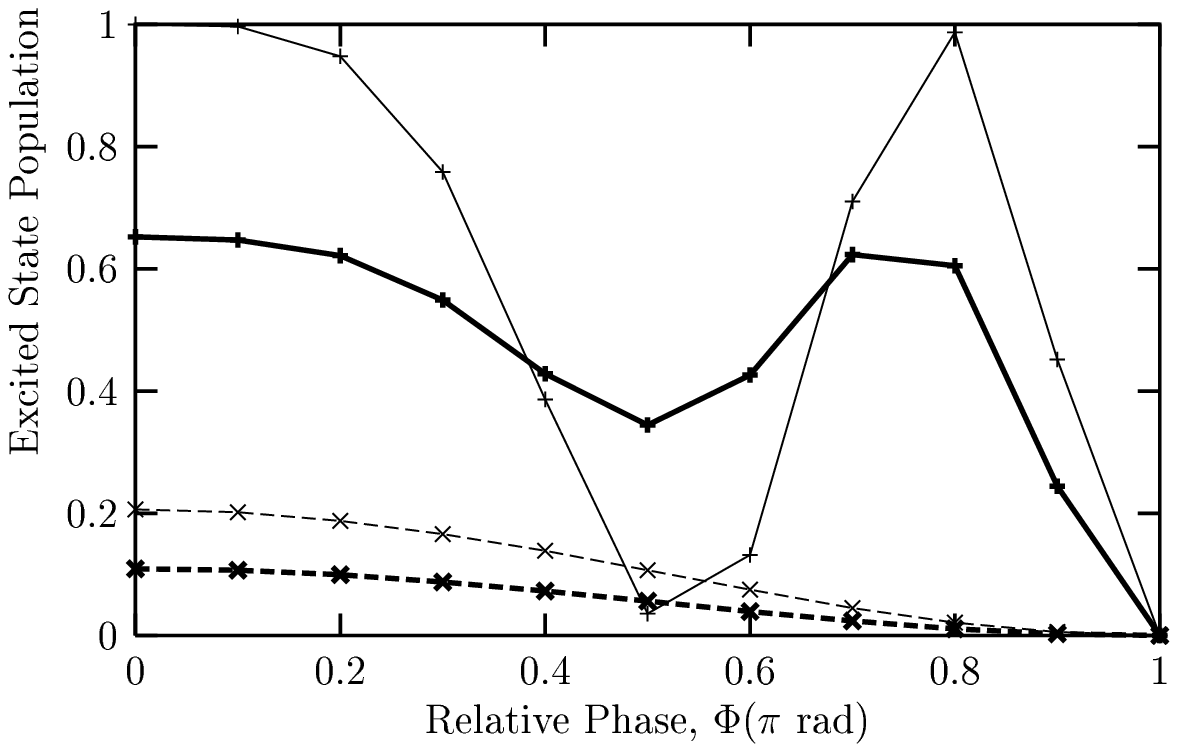}\\
   \includegraphics[angle=0,scale=1.]{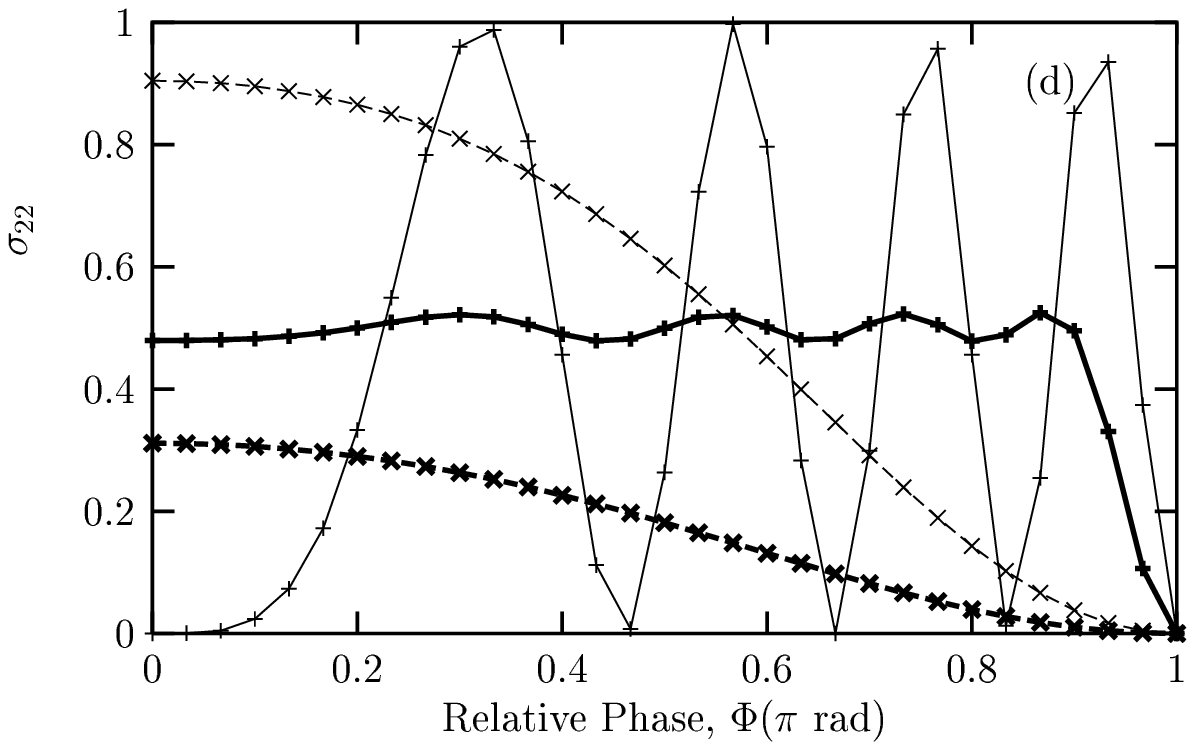}
\caption{}
   \label{control2lt}
  \end{center}\end{figure}

\end{document}